\documentclass[aps,pre,reprint,superscriptaddress,amsmath,amssymb,floatfix]{revtex4-2}

\usepackage{graphicx}
\usepackage{bm}
\usepackage{color}
\usepackage{hyperref}

\graphicspath{{figures/}{../figures/}}

\newcommand{\ii}{\mathrm{i}}

\begin{document}

\title{Self-propulsion in the 1D swarmalator model}

\author{K.~P.~O'Keeffe}
\affiliation{Starling Research Institute, Seattle, WA 98112, USA}

\date{\today}

\begin{abstract}
We study the 1D swarmalator model augmented with self-propulsion.  Each
swarmalator swims along the ring at a speed $v_0\sin\theta_i$ fixed by
its orientation $\theta_i$.  Self-propulsion unfolds the static states of
the ordinary model into traveling, breathing, split-wave, and chaotic
states.  Several of these states admit analytic reductions: an exact
drifting two-cluster branch with a closed-form stability spectrum, and a
four-cluster split-wave ansatz whose active pair reduces, in a
constant-orientation approximation, to an Adler equation.  Our numerical
evidence suggests that the transition to chaos under broad random initial
conditions is not caused by local destabilization of the ordered cluster
branches, but by basin reorganization among coexisting attractors.  The
resulting states may serve as qualitative signatures for confined active
oscillator arrays.
\end{abstract}

\maketitle

\section{Introduction}

Across biology and soft matter one finds groups of units that keep an
internal rhythm while they rearrange themselves in space, with each
process feeding the other: a unit's phase biases where it goes, and its
position decides whom it couples to.  Oscillators with this two-way
coupling between phase and motion are called \emph{swarmalators},
because they swarm and synchronize at the same time
\cite{OKeeffe2017Swarmalators}.  They turn up in settings as different as
swimming sperm~\cite{Riedl2023}, nematodes~\cite{Quillen2021},
magnetic domain walls~\cite{Hrabec2018}, colloidal
micromotors~\cite{Yan2012}, and robotic swarms~\cite{Barcis2019,Barcis2020}.

Since their introduction, swarmalators have been studied with
noise~\cite{Hong2023ThermalNoise}, external
forcing~\cite{Anwar2024Forced1D,Anwar2025ForcedHD},
pinning~\cite{Sar2023Pinning,Sar2024Driven,Sar2023Uncorrelated},
time delay~\cite{Blum2024Delay,OKeeffeHindes2026Delay},
attractive--repulsive interactions~\cite{Hao2023Attractive},
phase frustration~\cite{Lizarraga2023Sakaguchi},
distributed and finite-range
couplings~\cite{OKeeffeHong2022Distributed,Sar2025CouplingRange},
pulsating motion~\cite{Ghosh2025Pulsating,Yadav2025Pulse},
contrarians~\cite{Sar2025Contrarian},
higher-order interactions~\cite{Anwar2024HigherOrder},
topological structure~\cite{Gou2026Topological},
nonidentical frequencies~\cite{Yoon2022Nonidentical},
and in higher spatial dimensions~\cite{Yadav2024HigherDim},
together with general results on global synchronization and continuum
stability~\cite{OKeeffe2025Global,OKeeffe2025Stability}.

\begin{figure}[htbp]
  \centering
  \includegraphics[width=0.55\columnwidth]{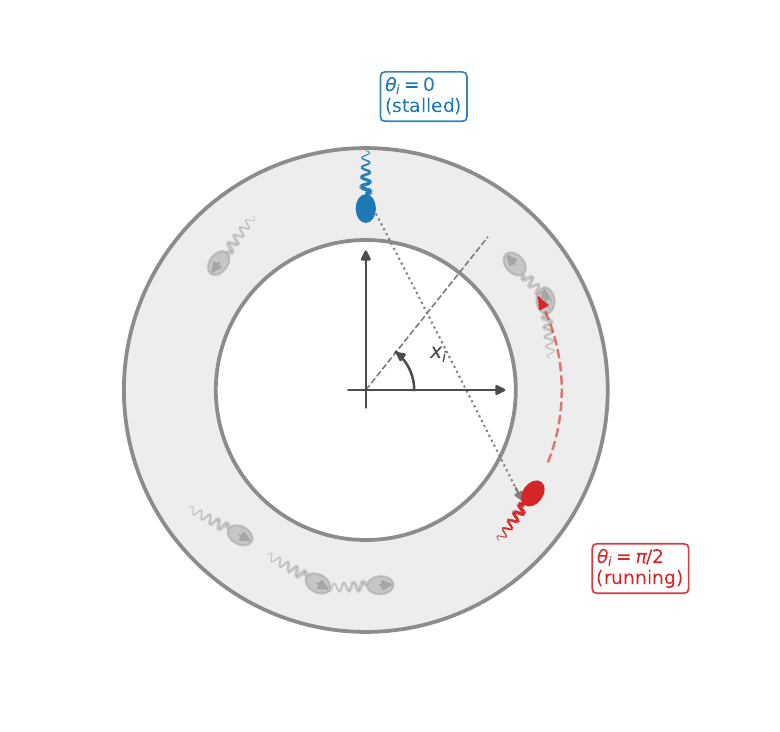}
  \caption{\textbf{Model schematic.}  Self-propelled swarmalators move
  on a 1D ring.  The orientation $\theta$ controls
  the tangential propulsion $v_0\sin\theta$.  The spatial interaction is
  attractive for similar orientations and repulsive for opposite ones,
  while the orientation coupling aligns nearby agents and anti-aligns
  agents on opposite sides of the ring.}
  \label{fig:1}
\end{figure}

We work with the one-dimensional version, confined to a ring
\cite{OKeeffe2022Ring,Yoon2022Nonidentical}.  Its appeal is that it is
simple enough to solve --- most of its states, \emph{async},
\emph{phase waves}, and \emph{sync}, can be written down in closed form
--- while still behaving like a genuine swarmalator system.  We add a
minimal self-propulsion: each agent moves at speed $v_0\sin\theta_i$,
treating $\theta_i$ as a body orientation rather than an abstract phase
(Fig.~\ref{fig:1}).  This places the model alongside active-matter
models of self-propelled and flocking particles
\cite{Vicsek1995,Czirok1999,Chate2008,Ginelli2010,Liebchen2017,Codina2022}.
The key structural difference from Vicsek-type models is the
bidirectional coupling between orientation and position: here the
spatial attraction is modulated by orientation similarity
\emph{and} the orientation alignment is modulated by spatial separation,
rather than orientation influencing position one-way.

We find that self-propulsion unfolds the static model into a multistable
dynamical system.  At small $v_0$, \emph{sync} deforms into a traveling
two-cluster state with an exact drift law and a closed-form linear
stability spectrum.  At intermediate $v_0$, a four-cluster \emph{split
wave} appears: two clusters sit where the propulsion vanishes, while two
sit where it is maximal.  In a constant-orientation four-cluster
reduction, the active-pair separation obeys an Adler equation, which
gives the breathing frequency and drift.
At larger $v_0$, a chaotic attractor appears and its basin grows; our
evidence suggests this is not a local destabilization of any ordered
state, but a reorganization of basins among coexisting attractors.
These states provide qualitative motifs --- drift, breathing, coexisting
cluster patterns --- for confined active oscillator arrays, though the
model is a minimal caricature rather than a model of any specific system.

\section{Model}
\label{sec:model}

\begin{figure*}[tp!]
  \centering
  \includegraphics[width=\textwidth]{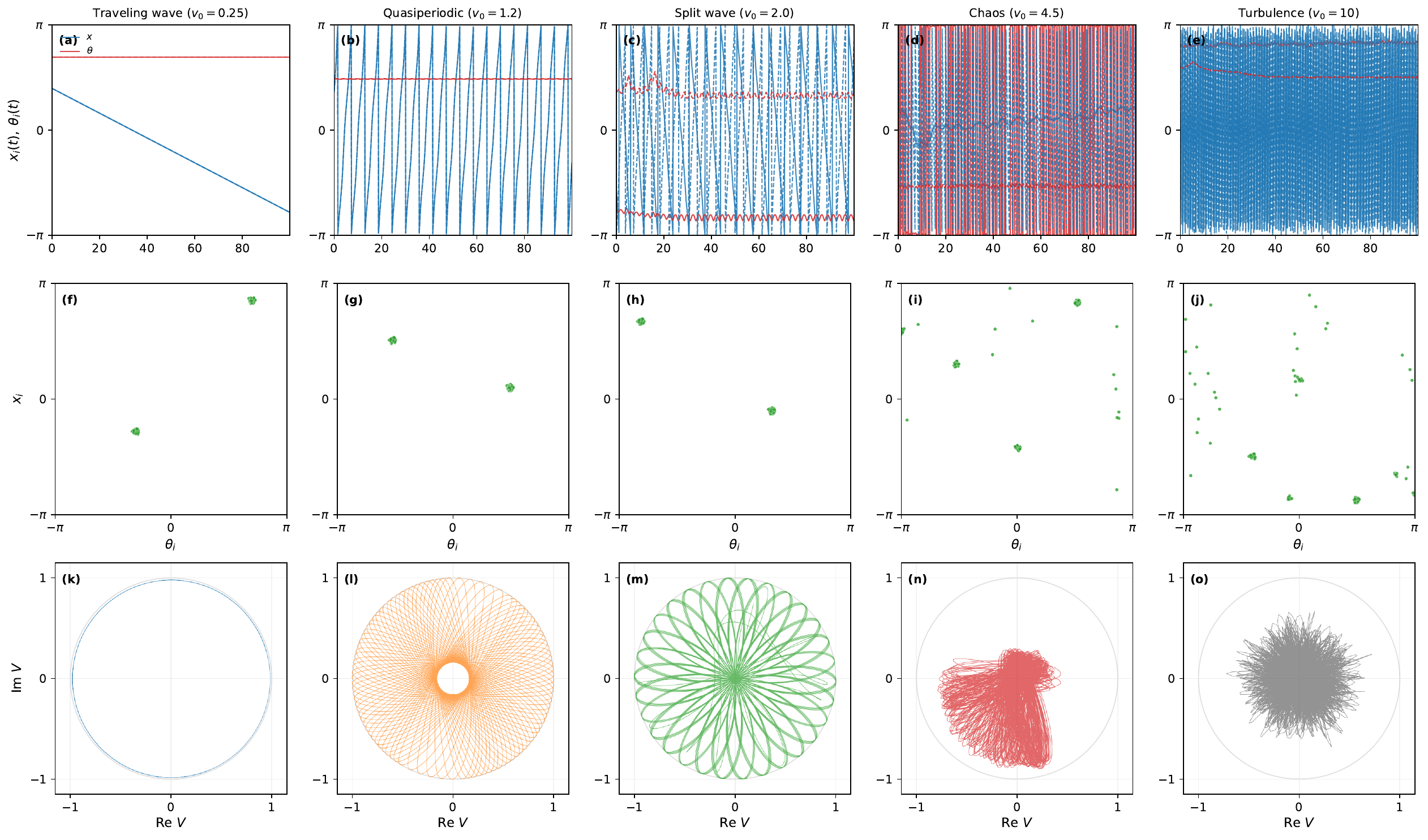}
  \caption{\textbf{Representative dynamics at $K=J=1$, $N=100$.}
  Five columns sample the cascade: traveling wave ($v_0=0.25$),
  breathing two-cluster ($v_0=1.2$), split wave
  ($v_0=2.0$), and chaos at two values ($v_0=4.5$ and $v_0=10$).
  Row 1: position $x_i(t)$ (blue) and orientation $\theta_i(t)$ (red)
  of two tagged particles (solid and dashed).
  Row 2: snapshot of all $N$ particles in the $(\theta_i, x_i)$ plane
  at the final time, with small jitter for visual contrast; the cluster
  count $m$ is read off directly.
  Row 3: trajectory of the order parameter $V(t)$ in the complex plane,
  with unit circle for reference.}
  \label{fig:2}
\end{figure*}

We consider $N$ identical self-propelled swarmalators on a ring of
circumference $2\pi$.  The position of agent $i$ is
$x_i\in[-\pi,\pi)$ and its orientation is
$\theta_i\in[-\pi,\pi)$.  The equations of motion are
\begin{align}
\dot x_i &= v_0\sin\theta_i
  + \frac{J}{N}\sum_{j=1}^{N}\sin(x_j-x_i)\cos(\theta_j-\theta_i),
  \label{eq:model_x}\\
\dot\theta_i &= \frac{K}{N}\sum_{j=1}^{N}\sin(\theta_j-\theta_i)
  \cos(x_j-x_i). \label{eq:model_theta}
\end{align}
The first term in Eq.~\eqref{eq:model_x} is self-propulsion projected
onto the tangent of the ring.  Thus $\theta_i=0$ corresponds to a radial
orientation and no tangential motion, while $\theta_i=\pi/2$ corresponds
to full counterclockwise motion with speed $v_0$.  The second term in
Eq.~\eqref{eq:model_x} is the usual spatial attraction of the 1D
swarmalator model, modulated by orientation similarity.  Equation
\eqref{eq:model_theta} aligns the orientations in a Kuramoto-like way,
modulated by spatial separation.  For $J>0$ we set $J=1$ by rescaling
time, so $K$ and $v_0$ are measured in units of $J$, and we vary
$(K,v_0)$.

When $v_0=0$, Eqs.~\eqref{eq:model_x}--\eqref{eq:model_theta} reduce to
the identical-frequency 1D swarmalator model.  In that limit the familiar
static states are: \emph{async}, with swarmalators uniformly distributed
in both $x$ and $\theta$; \emph{phase waves}, with
$x_i\pm\theta_i=\mathrm{const}$; and \emph{sync}, with all agents sharing
the same position and orientation
\cite{OKeeffe2022Ring,Yoon2022Nonidentical}.  The self-propulsion term
breaks the static character of these states while preserving the ring
symmetries.

We use the standard rainbow order parameters
\begin{align}
U &= r e^{\ii\phi} := \frac{1}{N}\sum_{j=1}^{N} e^{\ii(x_j+\theta_j)},
\nonumber\\
V &= s e^{\ii\psi} := \frac{1}{N}\sum_{j=1}^{N} e^{\ii(x_j-\theta_j)}.
\label{eq:rainbow}
\end{align}
The quantities $r=|U|$ and $s=|V|$ measure order in the two rainbow
coordinates $\xi=x+\theta$ and $\eta=x-\theta$.  We also compute the
maximum Lyapunov exponent $\lambda_{\max}$ by integrating the tangent
linear system and periodically renormalizing the perturbation vector.

\section{Numerics}
\label{sec:numerics}

Figure~\ref{fig:2} summarizes the representative
deterministic dynamics at $K=J=1$.  As $v_0$ is increased, four regimes appear in sequence
(Secs.~\ref{sec:clusters}--\ref{sec:multistability}).

\emph{Traveling cluster} (Fig.~\ref{fig:2}, column 1).
At small $v_0$ the $v_0=0$ sync state deforms into two
counter-oriented clusters that drift around the ring (under
broad random initial conditions; initial conditions restricted
to $\theta\in[0,\pi]$ instead produce a single traveling cluster).
The microscopic time series are linear ramps in $x$ at constant
$\theta$; the snapshot shows the two-cluster geometry; $V$ traces
a circle.

\emph{Breathing two-cluster motion}
(Fig.~\ref{fig:2}, column 2).  The two clusters begin to
breathe: the microscopic $\theta_i$ stays near constant while $x_i$
oscillates, the snapshot still shows two clusters, and $V$ fills a
Lissajous figure with two incommensurate frequencies.

\emph{Split wave} (Fig.~\ref{fig:2}, column 3;
Sec.~\ref{sec:splitwave}).  Four clusters appear, two at the
propulsion's zeros ($\theta\approx 0,\pi$) and two at the extrema
($\theta\approx\pm\pi/2$).  The cluster geometry --- visible as four
points in the row-2 snapshot --- distinguishes the split wave from
the breathing two-cluster, even though their $V$-plane Lissajous
patterns look superficially similar.

\emph{Chaos} (Fig.~\ref{fig:2}, columns
4--5).  At larger $v_0$ the microscopic dynamics become irregular,
the snapshot shows scattered particles, and $V$ wanders over a region
of the unit disk; $\lambda_{\max}$ becomes positive.  Two values of
$v_0$ are shown: at $v_0=10$ the cluster structure has fully dissolved
and $|V|$ stays small throughout.

\begin{figure}[htbp]
  \centering
  \includegraphics[width=\columnwidth]{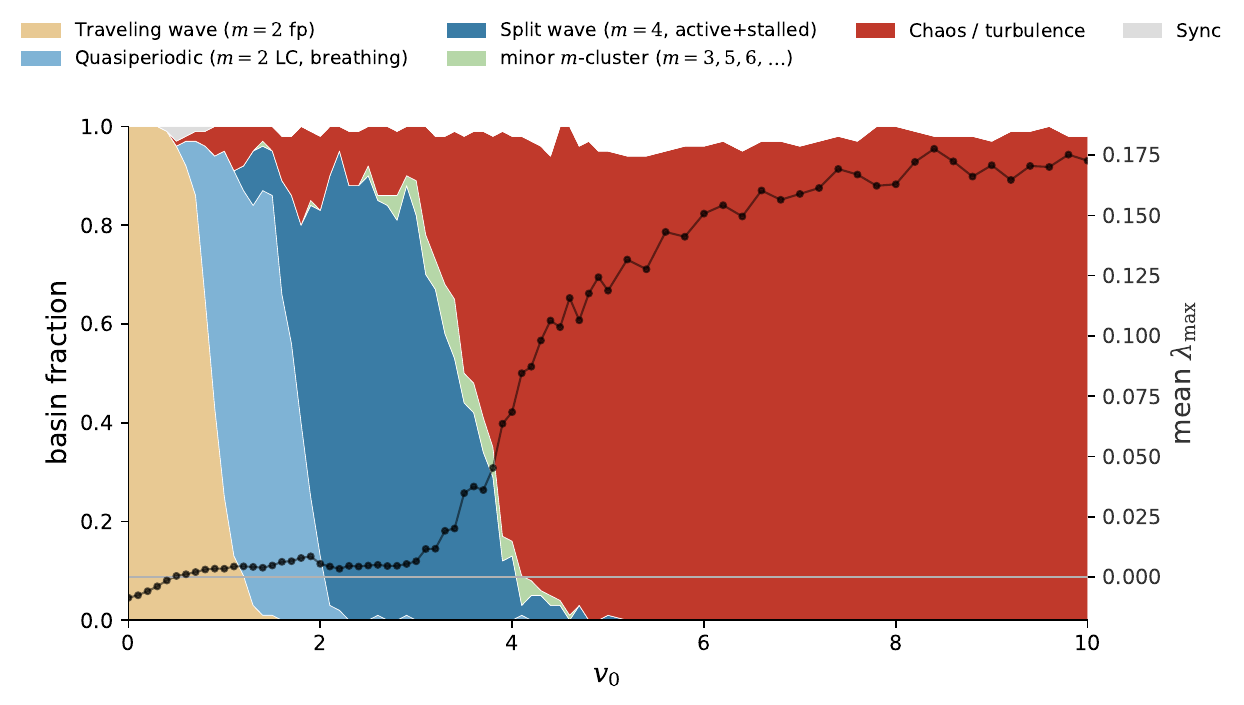}
  \caption{\textbf{Basin fractions at $K=J=1$.}
  Stacked-area basin fractions versus $v_0$ from $N=100$ direct
  integration with $100$ random uniform initial conditions per $v_0$
  ($T_{\rm trans}=200$, $T_{\rm record}=500$).  The dominant basin
  passes through four regimes --- traveling wave, breathing
  two-cluster, split wave, chaos --- with sync and minor
  $m$-cluster attractors as additional minority basins.  Black
  symbols (right axis): seed-averaged $\lambda_{\max}$.  These
  Lyapunov values are finite-step estimates; only their sign is used as
  a chaos diagnostic.}
  \label{fig:3}
\end{figure}

These states coexist as separate attractors over wide parameter
ranges; Fig.~\ref{fig:3} shows their basin fractions
versus $v_0$, discussed further in Sec.~\ref{sec:multistability}.

\section{Traveling two-cluster state}
\label{sec:clusters}

\textit{Existence.}---The low-$v_0$ ordered state can be understood analytically.  Suppose the
population splits into two clusters.  A fraction $p=n_1/N$ lies at
$(x_1,\theta_1)$, while the remaining fraction $q=1-p=n_2/N$ lies at
$(x_2,\theta_2)$, with
\begin{equation}
    x_2=x_1+\Delta x, \qquad \theta_2=\theta_1+\pi .
\end{equation}
The two clusters point in opposite directions.  Since
$\sin(\theta_2-\theta_1)=0$, the $\theta$ equation gives
$\dot\theta_1=\dot\theta_2=0$.  The spatial velocities are
\begin{align}
\dot x_1 &= v_0\sin\theta_1 - Jq\sin\Delta x,\\
\dot x_2 &= -v_0\sin\theta_1 + Jp\sin\Delta x.
\end{align}
A traveling two-cluster solution requires these two velocities to be
equal.  Therefore
\begin{equation}
    \sin\Delta x = \frac{2v_0}{J}\sin\theta_1^*.
    \label{eq:cluster_sep}
\end{equation}
The common drift speed is then
\begin{equation}
    \Omega = v_0(2p-1)\sin\theta_1^*.
    \label{eq:Omega}
\end{equation}
This formula has a simple interpretation.  If the two clusters are equal
in size, their active velocities cancel and the state does not drift.  If
one cluster is larger, the imbalance produces a net motion proportional
to the population difference $2p-1$.  Both the cluster orientation
$\theta_1^*$ and the population split $p$ are free parameters fixed by
the initial condition, not selected by the dynamics; the drift law is
therefore exact but conditional on these values.  The split $p$ is an
initial-condition-selected label of the cluster partition rather than a
dynamical degree of freedom: at finite $N$ it is discrete (a multiple of
$1/N$), and different $p$ give distinct two-cluster fixed points, not
points along a continuous neutral direction.  The genuine neutral modes
are the two zero eigenvalues found below (a global position rotation and
the branch orientation angle).

The branch exists whenever
\begin{equation}
    \left|\frac{2v_0}{J}\sin\theta_1^*\right|\le 1.
    \label{eq:existence}
\end{equation}
Thus the branch terminates for sufficiently large $v_0$, depending on
the IC-selected value of $\theta_1^*$, where the two solutions of
Eq.~\eqref{eq:cluster_sep} for $\Delta x$ merge ($|2v_0\sin\theta_1^*/J|
\to 1$, i.e.\ $C\to 0$ in the spectrum below).  As the spectrum shows,
two of the collective relaxation rates, $-KC$ and $-JC$, vanish at this
endpoint, consistent with a saddle-node/fold of the cluster geometry.
The endpoint is not a Hopf bifurcation: no complex-conjugate pair is
present anywhere on the branch.

The order parameters can also be written explicitly.  For the two-cluster
state,
\begin{align}
U &= e^{\ii(x_1+\theta_1)}\left[p-qe^{\ii\Delta x}\right],\\
V &= e^{\ii(x_1-\theta_1)}\left[p-qe^{\ii\Delta x}\right].
\end{align}
Thus $|U|=|V|$ and
\begin{equation}
    |U|^2=|V|^2=p^2+q^2-2pq\cos\Delta x.
    \label{eq:cluster_op}
\end{equation}
For $v_0=0$, the stable branch has $\Delta x=\pi$, so $|U|=|V|=1$.
Thus the state lies on the $r=s=1$ synchronous, or rainbow-synchronous,
manifold of the 1D ring model.  In the original $(x,\theta)$ variables
this point is represented as two antipodal clusters related by
$(x,\theta)\mapsto (x+\pi,\theta+\pi)$, which are equivalent in the
rainbow variables $(\xi,\eta)$.

\textit{Stability.}---The two-cluster branch is also linearly tractable.  Linearizing the full
$2N$-dimensional system around the two-cluster state gives a spectrum
that separates into inter-cluster modes and within-cluster spreading
modes.  Let
\begin{equation}
    C:=\sqrt{1-\left(\frac{2v_0}{J}\sin\theta_1^*\right)^2},
\end{equation}
where $C=-\cos\Delta x$ on the branch connected continuously to the
$v_0=0$ synchronized state.  The eigenvalues are
\begin{align}
\lambda_{1,2} &= 0 \quad (\times 2), \nonumber\\
\lambda_{3} &= -KC, \quad \lambda_{4} = -JC, \nonumber\\
\lambda_{5} &= -J(p+qC) \quad (\times\, n_1-1), \nonumber\\
\lambda_{6} &= -K(p+qC) \quad (\times\, n_1-1), \nonumber\\
\lambda_{7} &= -J(q+pC) \quad (\times\, n_2-1), \nonumber\\
\lambda_{8} &= -K(q+pC) \quad (\times\, n_2-1).
\label{eq:spectrum}
\end{align}

Every nonzero eigenvalue in Eq.~\eqref{eq:spectrum} is negative for all
$v_0$ (with $0<C\le 1$ and $K,J>0$): the two-cluster state never loses
linear stability while the branch exists.  This is itself informative.
It means the state's disappearance from the dominant basin at moderate
$v_0$ (Sec.~\ref{sec:multistability}), where the branch is robustly
stable and the eigenvalues remain strictly negative, cannot be a local
instability of this branch and must instead be a global (basin) effect,
consistent with the
multistability picture below.  The spectrum also suggests which cluster
is most weakly damped.  Among the intra-cluster spreading modes, the minority
cluster (factor $p+qC$) is less strongly damped than the majority
cluster ($q+pC$) for $p<q$, $0<C<1$, so the first width oscillations
appear on the minority cluster.  (The globally slowest modes are in fact
the collective ones $-KC$ and $-JC$, since $C<p+qC$ for $C<1$; but these
are rigid translations of the cluster centroids, not changes in cluster
width.)  This suggests a microscopic interpretation of where the first
unsteady motion appears as the breathing branch is approached.

\section{Split-wave state}
\label{sec:splitwave}

\begin{figure}[htbp]
  \centering
  \includegraphics[width=0.9\columnwidth]{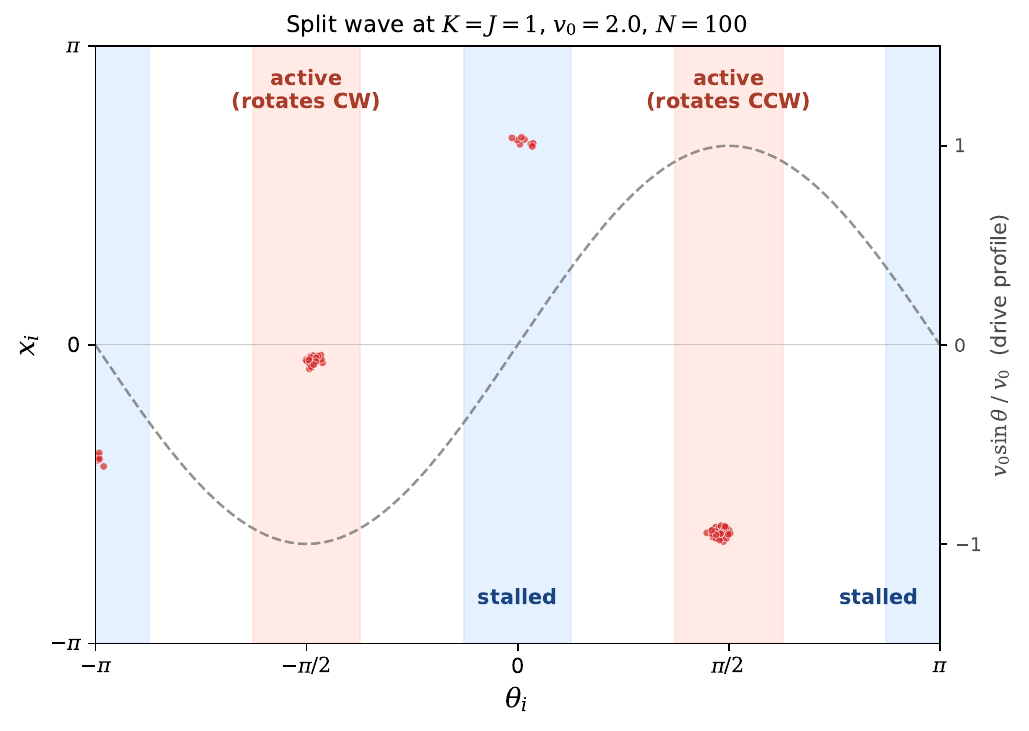}
  \caption{\textbf{Split-wave geometry.}  Particles in the
  $(\theta,x)$ plane from a representative split-wave attractor at
  $K=J=1$, $v_0=2$, $N=100$.  Stalled clusters at
  $\theta\approx 0,\pi$ (zero propulsion); active clusters at
  $\theta\approx\pm\pi/2$ (maximum propulsion, opposite directions).
  Particles are jittered slightly for visual contrast; the
  underlying clusters are nearly delta-function tight in
  $(\theta,x)$.}
  \label{fig:4}
\end{figure}

Above $v_0\approx 1.8$ at $K=J=1$, the dominant attractor
is no longer the two-cluster traveling state but a new four-cluster
state.  Inspection of representative snapshots
(Fig.~\ref{fig:4}) shows that the cluster centers
sit close to the four special points of the active drive
$v_0\sin\theta$:
\begin{itemize}
\item Two \emph{stalled} clusters at $\theta\approx 0$ and $\theta\approx\pi$,
where $\sin\theta=0$ and the propulsion vanishes;
\item Two \emph{active} clusters at $\theta\approx +\pi/2$ and
$\theta\approx -\pi/2$, where $|\sin\theta|=1$ and the propulsion is
maximal in opposite directions.
\end{itemize}
We verified this geometric pinning across $1139$ four-cluster snapshots
drawn from $48$ random initial conditions at each of six values
$v_0\in[1.8,3.5]$ ($K=J=1$, $N=100$).  Measuring for each snapshot the
largest deviation of any cluster center from the canonical
$\{0,\pm\pi/2,\pi\}$ positions, the median deviation is $8.6^\circ$, and
the worst-cluster deviation is within $15^\circ$ in $80\%$ of snapshots,
within $20^\circ$ in $92\%$, and within $25^\circ$ in $96\%$ (the small
tail up to $\sim\!45^\circ$ is sampled during the limit-cycle
breathing).  Among the simplest $m=4$ configurations, the
$\pi/2$-spaced one lets the cluster orientations form two antipodal
pairs simultaneously aligned with the propulsion's zeros and its
extrema; this maximizes the antipodal-pair-cancellation symmetry
that we exploit analytically below.  We refer to this state as a
\emph{split wave}: a fast-rotating active subpopulation (clusters
at $\theta=\pm\pi/2$) coupled to a stalled subpopulation (clusters
at $\theta=0,\pi$ where the propulsion vanishes), with positions
dictated by the propulsion's special points.  We have verified the
geometric pinning at $K=J=1$ only; whether the same $\pi/2$-spaced
configuration persists at general $K>0$, or whether the cluster
positions deform with $K$, is left to future work.

\textit{Approximate reduction.}---We do not derive the split wave from
scratch; instead we take the $\pi/2$-spaced geometry as an ansatz
motivated by the numerical pinning above and ask what dynamics it
implies.  We approximate the population as four clusters with fractions $p_k$ ($\sum p_k = 1$) and
orientations held at $\Theta_k=\pi k/2$, $k=0,1,2,3$ (the constant-$\Theta_k$
approximation, exact only as $K\to 0$; see below).  Substituting into
Eqs.~\eqref{eq:model_x}--\eqref{eq:model_theta} and using
$\cos(\Theta_l-\Theta_k)=\cos(\pi(l-k)/2)\in\{1,0,-1,0\}$, the
position equation for cluster $k$ involves only its antipodal partner
$k+2\pmod 4$:
\begin{equation}
\dot X_k = v_0\sin\Theta_k - J\,p_{k+2}\sin(X_{k+2}-X_k).
\label{eq:m4_pos}
\end{equation}
The four cluster-centroid coordinates decouple into two antipodal
pairs: $\{X_0,X_2\}$ at the propulsion zeros (the stalled pair) and
$\{X_1,X_3\}$ at the propulsion extrema (the active pair).

\textit{Stalled pair.}---With no $v_0$ forcing, the separation
$D_{02}\equiv X_2-X_0$ obeys
\begin{equation}
\dot D_{02} = J(p_0+p_2)\sin D_{02},
\end{equation}
with stable fixed point $D_{02}=\pi$.  The stalled clusters lock at
antipodal $x$-separation throughout the dynamics.

\textit{Active pair: Adler equation.}---For the active pair, the
separation $D_{13}\equiv X_3-X_1$ obeys
\begin{equation}
\dot D_{13} = -2v_0 + J(p_1+p_3)\sin D_{13}.
\label{eq:m4_adler}
\end{equation}
This is the classical Adler equation~\cite{Adler1946,Strogatz_book}
with detuning $\Delta=-2v_0$ and
coupling $b=J(p_1+p_3)$.  It admits a stable fixed point (locked
$D_{13}$) for
\begin{equation}
v_0 \le \frac{J(p_1+p_3)}{2},
\label{eq:m4_fold}
\end{equation}
and is unlocked above this saddle-node fold, with $D_{13}$ wrapping
periodically at frequency
$\Omega_D=\sqrt{4v_0^2-J^2(p_1+p_3)^2}$.
The fold value $J(p_1+p_3)/2$ is the prediction of the constant-$\Theta_k$
reduction.  At finite $K$ the cluster orientations are not exactly
pinned, and the empirical limit-cycle birth in the full collective-ODE
dynamics is shifted upward substantially: for equal clusters it moves
from $J/4$ to $v_0\approx1.3$ at $K=J=1$, and numerically this shift
shrinks as $K/v_0$ decreases, recovering $J/4$ as $K\to0$.

\begin{figure}[htbp]
  \centering
  \includegraphics[width=\columnwidth]{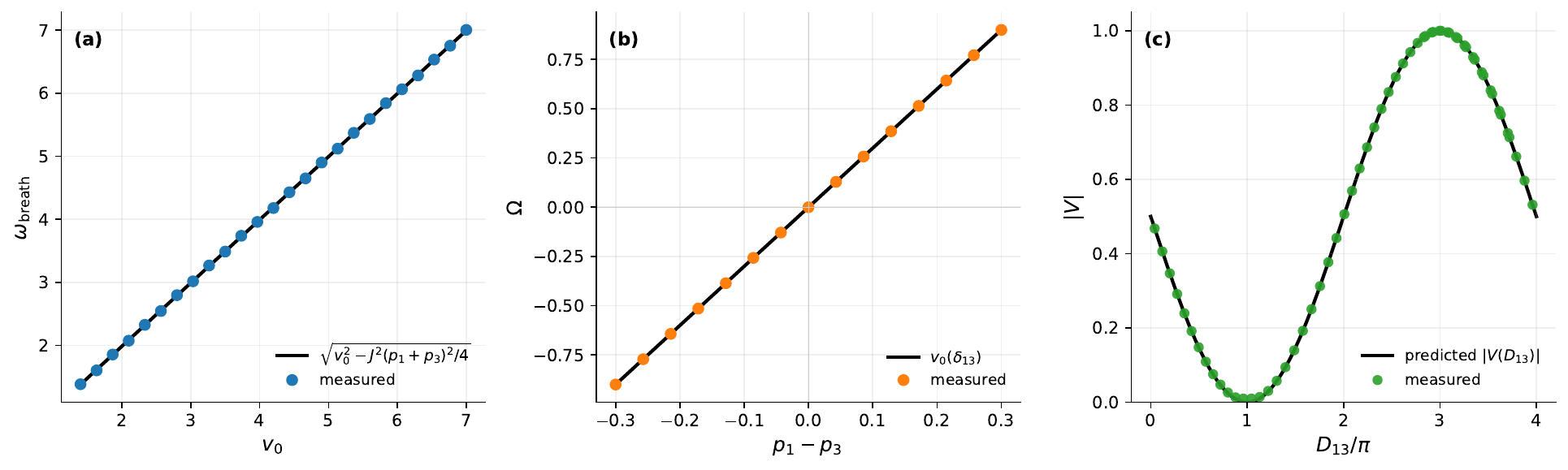}
  \caption{\textbf{Internal check of the constant-$\Theta_k$ reduced-ODE
  predictions ($J=1$).}
  (a)~Breathing frequency $\omega_{\rm breath}$ vs $v_0$ for the
  symmetric variant ($p_1=p_3=1/4$): predicted
  $\sqrt{v_0^2-J^2/16}$ (line) vs measured FFT peak (circles).
  (b)~Centroid drift $\Omega$ vs within-pair imbalance $p_1-p_3$
  at $v_0=3$: predicted $v_0(p_1-p_3)$ (line) vs measured
  centroid drift (circles).
  (c)~Gauge-invariant order-parameter amplitude $|V|$ vs the active-pair
  separation $D_{13}$ for the symmetric split wave ($p_k=1/4$, $v_0=2.5$):
  measured (green) collapses onto the prediction $|V(D_{13})|$ from
  the $\delta=0$ specialization of Eq.~\eqref{eq:m4_V_full} (black).}
  \label{fig:5}
\end{figure}

\textit{Order parameter.}---Substituting the four-cluster ansatz into
$V$, with $X_2=X_0+\pi$, $M=(X_1+X_3)/2$,
$D=D_{13}=X_3-X_1$, and $\delta=M-X_0$, gives
\begin{equation}
V=e^{\ii X_0}\left[
S_0+e^{\ii\delta}
\left(-S_1\sin(D/2)-\ii\Delta p\cos(D/2)\right)
\right],
\label{eq:m4_V_full}
\end{equation}
where $S_0=p_0+p_2$, $S_1=p_1+p_3$, and
$\Delta p=p_1-p_3$.  Consequently
\begin{align}
|V|^2 &= S_0^2+S_1^2\sin^2(D/2)+\Delta p^2\cos^2(D/2)
\nonumber\\
&\quad -2S_0S_1\sin(D/2)\cos\delta
  +2S_0\Delta p\cos(D/2)\sin\delta .
\label{eq:m4_V_amp}
\end{align}
Thus the relative phase between the stalled pair and the active-pair
midpoint is not a removable global phase in general.  In the
constant-$\Theta_k$ reduction this relative phase contains a neutral
offset set by the initial condition (and, for imbalanced active pairs,
can evolve with $D$); in the finite-$K$ full system it may be selected
dynamically.  The symmetric-frame specialization $\delta=0$ gives
\begin{equation}
V = (p_0+p_2) - (p_1+p_3)\sin(D_{13}/2) - \mathrm{i}(p_1-p_3)\cos(D_{13}/2).
\label{eq:m4_V}
\end{equation}
For the symmetric wave with this frame choice, the $|V|$ breathing
prediction is parameter-free and matches the reduced ODE
(Fig.~\ref{fig:5}); its frequency follows from the
rate at which $|V|$ completes a cycle,
\begin{equation}
\omega_{\rm breath} = \tfrac{1}{2}\Omega_D = \sqrt{v_0^2 - J^2(p_1+p_3)^2/4}.
\label{eq:m4_freq}
\end{equation}

\textit{Drift law.}---The total population-weighted centroid drift is
\begin{equation}
\Omega = v_0(p_1-p_3),
\label{eq:m4_drift}
\end{equation}
arising entirely from the active pair's asymmetry; the stalled pair
contributes zero by symmetry.  The split wave thus has two sub-branches:
a non-drifting variant ($p_1=p_3$, $\Omega=0$) and a drifting one
($p_1\ne p_3$, $\Omega\ne 0$).

\textit{Family of partitions.}---The fractions $p_k$ are not fixed by the
cluster equations; like $p$ in the two-cluster state, they are
initial-condition-selected labels of the cluster partition rather than
dynamical variables.  The split wave therefore comes as a two-parameter
family of partitions, labeled by the within-pair imbalances
$\delta_{02}=p_0-p_2$ and $\delta_{13}=p_1-p_3$ at fixed pair sums.
This generalizes the single-parameter population split $p$ of the
two-cluster traveling state (Sec.~\ref{sec:clusters}).

Two parts of the reduction are checked directly against the full $N=100$
system: the geometric pinning (the $1139$-snapshot statistic above), and
stability --- a split-wave ansatz with intra-cluster noise $\sigma=0.01$
at $v_0\in\{4,5,6\}$ relaxes to zero cluster width.  The predicted
$\omega_{\rm breath}$, $\Omega$, and $|V|$ breathing are verified against
the weighted four-cluster collective ODE
(Fig.~\ref{fig:5}).  This is an internal check of
the constant-$\Theta_k$ reduction: since the approximation is exact only
as $K\to0$, the figure confirms the reduction's self-consistency but does
not verify that the predicted formulas match the finite-$K$ full-$N$
body breathing quantitatively.

\section{Multistability and chaos}
\label{sec:multistability}

Figure~\ref{fig:3} shows the basin fractions at $K=J=1$.
The dominant basin transitions through four regimes: traveling wave
($v_0\lesssim 0.8$), breathing two-cluster ($v_0\in[0.8,1.8]$), split
wave ($v_0\in[1.8,3.5]$), and chaos ($v_0\gtrsim 3.5$).

The key feature is that attractor existence and basin dominance are
decoupled.  At $v_0=2$ (split-wave-dominant), 10/10 initial conditions
perturbed from the two-cluster ansatz remain on the two-cluster family,
and 10/10 perturbed-split-wave initial conditions remain on the split
wave, even though random ICs find the two-cluster family only
$\sim\!13\%$ of the time.  At $v_0=4.5$ (chaos-dominant), both cluster
attractors still persist as stable but vanishing-basin objects.  This
separation is the signature of the multistable cascade.

The synchronized state ($m=1$) is the degenerate $p\to 1$ limit of the
two-cluster spectrum; it is linearly stable for all $v_0$ but its basin
under broad random initial conditions is small.

At $v_0\gtrsim 3.5$ a chaotic attractor of the finite-$N$ system takes
over the random-IC basin.
It is qualitatively distinct from the cluster states: its maximal
Lyapunov exponent is positive
(Fig.~\ref{fig:A1}; robust to integration time, step size,
and classifier threshold, Appendix~\ref{app:robustness}), and it has no
well-defined cluster count --- gap clustering of instantaneous snapshots
at $v_0=7$ returns anywhere from a few to $\sim\!20$ apparent clusters.
Its finite-size dependence and continuum limit are left to future work.

This chaos is not a local destabilization of either cluster family;
both remain stable across the chaos window.  The two-cluster spectrum is
all-real (Eq.~\eqref{eq:spectrum}), and direct integration of perturbed
split-wave collective initial conditions stays on the eight-dimensional
collective orbit from $v_0=2$ to $8$, while a perturbed split wave in the
full $N=100$ system ($\sigma=0.01$, $v_0\in\{4,5,6\}$) relaxes back
rather than spreading.  We therefore read
the chaos as a coexisting attractor whose basin is disjoint from the
cluster basins.  The MLE of the chaotic trajectories grows smoothly
from small positive values near $v_0\approx 2.5$ to $\approx 0.12$ at
$v_0\approx 5$, with no abrupt jump, which is consistent with a
boundary-crisis scenario in which the chaotic attractor exists at small
basin volume for moderate $v_0$ and grows as basin boundaries shift;
however, identifying the bifurcation precisely is left open.

The cascade is not special to $K=J=1$.  Representative sweeps at
$K=0.5,1.0,1.5$ ($25$ random initial conditions per cell, same
classifier) find the same four-regime sequence --- traveling wave,
breathing two-cluster, split wave, chaos --- with boundaries that rise
with $K$:
\begin{table}[h]
\centering
\caption{Approximate $v_0$ boundaries between cascade regimes at three
values of $K$ ($J=1$, $N=100$, $25$ random ICs per cell).}
\begin{tabular}{cccc}
\hline
$K$ & trav.$\to$breath. & breath.$\to$split & split$\to$chaos \\
\hline
$0.5$ & $0.6$ & $1.0$ & $2.0$ \\
$1.0$ & $0.8$ & $1.6$ & $3.0$ \\
$1.5$ & $1.0$ & $2.0$ & $3.5$ \\
\hline
\end{tabular}
\label{tab:boundaries}
\end{table}
\noindent (approximate $v_0$ boundaries; the coarse sweep, with shorter
integration, places them somewhat below the high-fidelity $K=1$ values
above, but the ordering and the upward trend are robust).  For larger
$K$ the split-wave window closes (near $K\approx 2$), and the breathing
two-cluster passes directly to chaos.  At $K<0$ the cascade is
structurally different again: a sweep at $K=-0.5$ finds no split wave and
chaos several times earlier.  The collective states and their ordering
are therefore robust over a substantial part of the aligning regime, not
artifacts of the single coupling at which the analytics were developed.

\section{Discussion}
\label{sec:discussion}

We have studied the 1D swarmalator model augmented with a minimal
self-propulsion term $v_0\sin\theta_i$.  The main finding is that this
single term converts a static, analytically tractable model into a
multistable dynamical system, and that much of the resulting attractor
landscape is itself analytically tractable.

Two cluster families admit closed-form reductions.  The
traveling two-cluster state has an exact drift law
$\Omega=v_0(2p-1)\sin\theta_1^*$ and a closed-form $2N$-dimensional
stability spectrum with all real eigenvalues; the all-real spectrum
rules out a Hopf bifurcation of the rigid branch, so the breathing
branch must arise via a global mechanism.  The four-cluster split wave
reduces, in a constant-orientation approximation, to an Adler equation
for the active-pair separation, giving closed-form breathing frequency,
drift law, and saddle-node fold.  Both families persist as stable
attractors across the entire cascade; the dominant basin under random
initial conditions reorganizes among them and eventually passes to a
chaotic attractor at large $v_0$.  Multistability, rather than a
sequence of local bifurcations, is the organizing principle.

In the pinned and driven swarmalator models, chaos appears via
period-doubling and intermittency routes from forced periodic
states~\cite{Sar2023Pinning,Sar2024Driven}.  Here, by contrast, chaos
arises from a self-propelled cascade among freely coexisting attractors;
the cluster families are not destroyed but simply lose basin share.
This mechanism is structurally different, and its identification suggests
that the swarmalator chaos taxonomy is richer than previously recognized.

The model suggests qualitative signatures for confined active oscillator
arrays: coherent circulation, two-frequency breathing, coexisting two-
and four-cluster patterns, and stationary subpopulations pinned at the
propulsion's zeros.  The connection to bacterial or sperm racetrack
experiments~\cite{Wioland2016,Creppy2016} is qualitative; the model is
a minimal caricature, not a fit to any specific system.

Several questions remain open.  The birth mechanism of the chaotic
attractor deserves further study; the smooth onset of the maximal
Lyapunov exponent is consistent with a boundary crisis, but a definitive
identification is lacking.  Whether the chaotic attractor survives the
continuum limit $N\to\infty$ is also open, given the observed $N$-scaling
of $\lambda_{\max}$.  Natural extensions include the full $(K,v_0)$
phase diagram, analytics for higher-$m$ minority cluster attractors,
and models with distributed natural frequencies or more realistic swimmer
orientations.

\paragraph*{Data and code availability.}
The simulation code, raw trajectory data, and figure-generation scripts
used to produce all results in this paper will be made available in a
public repository upon acceptance.

\begin{acknowledgments}
The author thanks collaborators and colleagues in the swarmalator and
active-matter communities for useful discussions.
\end{acknowledgments}

\appendix

\section{Numerical methods}
\label{app:numerics}

Unless otherwise stated, simulations used $J=1$ and random initial
conditions with $x_i,\theta_i$ independently and uniformly distributed on
$[-\pi,\pi)$.  Time integration used a fixed-step fourth-order
Runge--Kutta method with step $\Delta t=0.1$, with $x_i$ and $\theta_i$
wrapped to $[-\pi,\pi)$ at every step.  Basin fractions
(Fig.~\ref{fig:3}) discarded a transient $T_{\rm trans}=200$
and recorded over $T_{\rm record}=500$.  The maximum Lyapunov exponent
was computed by co-integrating the tangent linear system associated with
Eqs.~\eqref{eq:model_x}--\eqref{eq:model_theta}, renormalizing the
perturbation vector every $20$ steps, over $T_{\rm MLE}=2000$ after a
transient $T_{\rm trans}=300$; as the step size enters its magnitude
(Appendix~\ref{app:robustness}), we use only its sign.

The two-cluster branch and the synchronized state coexist with the
unsteady attractors at every $v_0$.  The cascade reported in
Sec.~\ref{sec:multistability} is observed in the basin selected by
broad random initial conditions described above; narrower initial
distributions in $\theta$ instead converge to a synchronized rotating
branch and yield non-positive maximum Lyapunov exponents.
Stability of the eight-dimensional split-wave collective orbit was
checked by direct integration from perturbed collective initial
conditions.

\textit{Attractor classifier.}---The basin-fraction tallies in
Sec.~\ref{sec:multistability} use an automated classifier
applied to each trajectory's final-state and order-parameter
diagnostics.  After discarding transients, the classifier (i)
computes the cluster count $m$ by gap clustering of $\theta_{\rm final}$
--- reliable for the ordered cluster states studied here, and checked
against the full $(\theta,x)$ snapshots for representative cases;
(ii) detects chaos as $\lambda_{\max}>0.01$; (iii)
distinguishes fixed points from limit cycles via $\mathrm{std}(|V|)$;
(iv) distinguishes drifting limit cycles (quasiperiodic in $V$) from
non-drifting ones (periodic in $V$) via the Hermitian-symmetry score
of the two-sided FFT of the complex order parameter $V(t)$.  Let
$F_k=\bigl[\mathrm{FFT}(V-\langle V\rangle)\bigr]_k$, with positive
indices $k>0$ corresponding to positive frequencies and negative
indices $k<0$ to negative frequencies; the score is
$H=2\sum_{k>0}\min(|F_k|,|F_{-k}|)/\sum_{k>0}(|F_k|+|F_{-k}|)$.
For a real-valued $V$ (no drift) the spectrum is Hermitian-symmetric
and $H\to 1$; for a drifting $V$ the spectrum is asymmetric and
$H<0.7$.  We caution that the limit-cycle/quasiperiodic boundary is the
least sharp of these cuts: the breathing two-cluster state is formally a
two-frequency (drift plus breathing) motion, but when the breathing
sideband is weak the classifier labels it a limit cycle.  The
``quasiperiodic'' label should therefore be read as a description of the
underlying two-frequency motion rather than as a precise basin count;
none of the conclusions of Sec.~\ref{sec:multistability} depend on the
limit-cycle/quasiperiodic distinction.

The tangent equations are obtained by differentiating the vector field
with respect to $(x_i,\theta_i)$.  For example, the derivatives of the
$x$ equation include
\begin{align}
\frac{\partial \dot x_i}{\partial x_k}
&= \frac{J}{N}\cos(x_k-x_i)\cos(\theta_k-\theta_i), \qquad k\ne i,\\
\frac{\partial \dot x_i}{\partial \theta_k}
&= -\frac{J}{N}\sin(x_k-x_i)\sin(\theta_k-\theta_i), \qquad k\ne i,
\end{align}
with diagonal entries fixed by summing the corresponding negative
off-diagonal interaction terms and adding
$\partial(v_0\sin\theta_i)/\partial\theta_i=v_0\cos\theta_i$ to the
$x_i$--$\theta_i$ block.  The derivatives of the $\theta$ equation are
computed analogously.

\section{Derivation of the two-cluster branch}
\label{app:cluster_derivation}

For completeness, we derive Eq.~\eqref{eq:Omega}.  Assume two clusters
with fractions $p$ and $q=1-p$:
\begin{equation}
(x_2,\theta_2)=(x_1+\Delta x,\theta_1+\pi).
\end{equation}
The orientation velocities vanish because
$\sin(\theta_2-\theta_1)=\sin\pi=0$.  The spatial velocities are
\begin{align}
\dot x_1 &= v_0\sin\theta_1
  + Jq\sin(\Delta x)\cos\pi
  = v_0\sin\theta_1-Jq\sin\Delta x,\\
\dot x_2 &= v_0\sin(\theta_1+\pi) + Jp\sin(-\Delta x)\cos(-\pi)
  \nonumber\\
  &= -v_0\sin\theta_1+Jp\sin\Delta x.
\end{align}
Setting $\dot x_1=\dot x_2$ gives
\begin{equation}
2v_0\sin\theta_1=J\sin\Delta x,
\end{equation}
which is Eq.~\eqref{eq:cluster_sep}.  Substituting this relation into
$\dot x_1$ gives
\begin{equation}
\Omega=\dot x_1=v_0\sin\theta_1-2qv_0\sin\theta_1
      =v_0(2p-1)\sin\theta_1.
\end{equation}

\section{Notes on the stability spectrum}
\label{app:spectrum}

The spectrum in Eq.~\eqref{eq:spectrum} follows from the block structure
of the two-cluster Jacobian.  Perturbations decompose into four
collective directions, which move the two cluster centroids relative to
one another, and $2(n_1-1)+2(n_2-1)$ spreading directions, which split
agents inside either cluster without changing the corresponding cluster
mean.  This decomposition is the same finite-$N$ mechanism that makes
synchronized clusters analytically tractable in the ordinary 1D
swarmalator model.

The two neutral eigenvalues correspond to global position rotation and
the continuous orientation parameter $\theta_1^*$ labeling the
two-cluster branch.  The two collective stable eigenvalues are $-KC$ and
$-JC$.  The cluster-spread eigenvalues are
proportional to the effective attraction felt by perturbations within the
cluster.  For cluster 1 this factor is $p+qC$; for cluster 2 it is
$q+pC$.  These factors reduce to one at $v_0=0$, recovering the
synchronized-state damping rates of the static model.

\section{Robustness of the chaos diagnostics}
\label{app:robustness}

\begin{figure*}[t]
  \centering
  \includegraphics[width=0.85\textwidth]{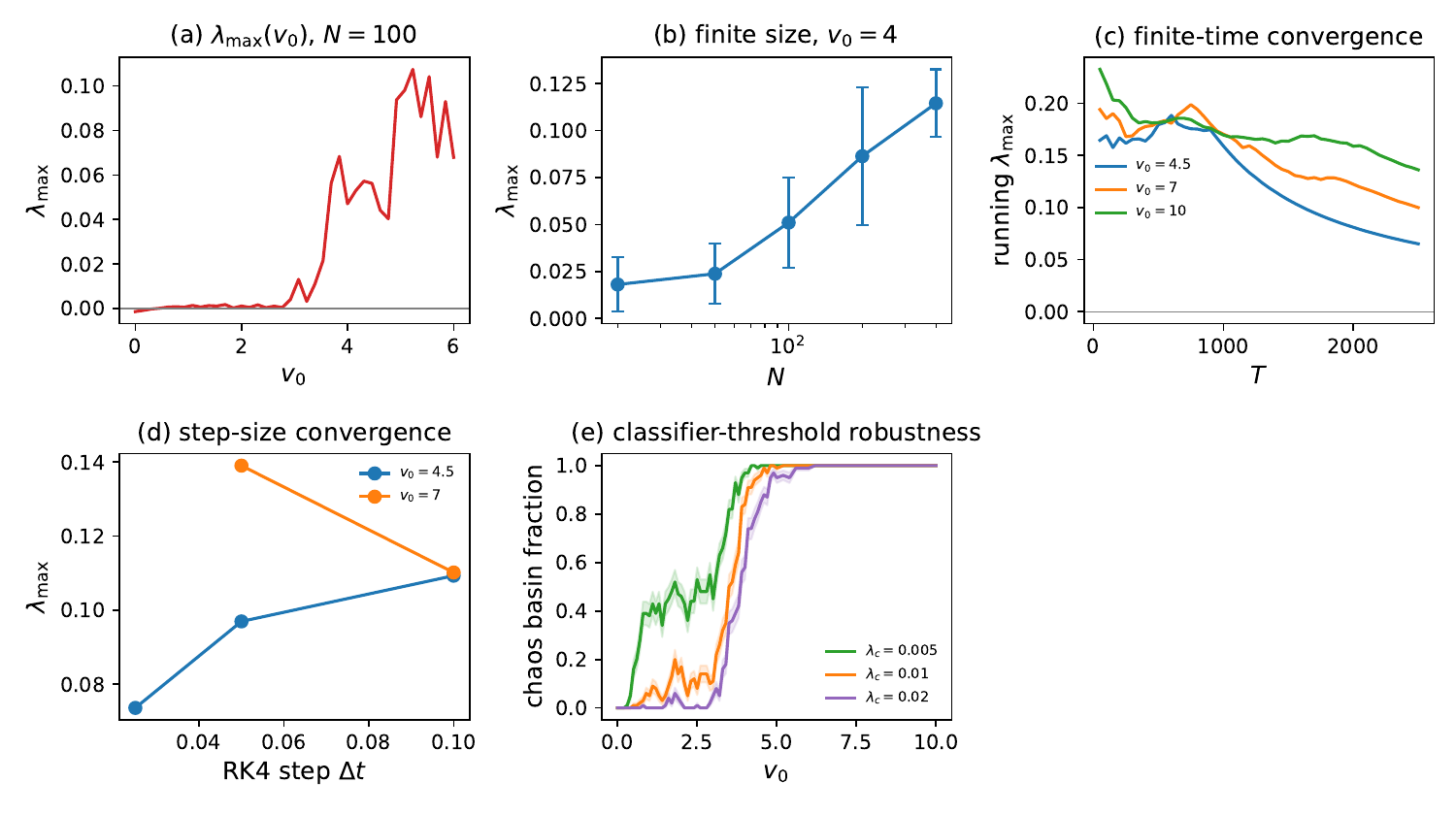}
  \caption{\textbf{Robustness of the chaos diagnostics at $K=J=1$.}
  (a) $\lambda_{\max}(v_0)$ at $N=100$;
  (b) finite-size dependence of $\lambda_{\max}$ at $v_0=4$;
  (c) finite-time convergence of the running $\lambda_{\max}(T)$;
  (d) RK4 step-size convergence of $\lambda_{\max}$;
  (e) chaos basin fraction versus $v_0$ for classifier thresholds
  $\lambda_c\in\{0.005,0.01,0.02\}$, with binomial uncertainty bands.}
  \label{fig:A1}
\end{figure*}

The chaotic attractor is diagnosed numerically, so we checked that the
diagnostics do not depend on the numerical controls
(Fig.~\ref{fig:A1}).  Panel (a): the maximal Lyapunov
exponent $\lambda_{\max}(v_0)$ at $N=100$ is non-positive in the ordered
windows and rises smoothly to positive values in the chaos window.
Panel (b): at fixed $v_0=4$, $\lambda_{\max}$ grows with system size over
$N=20$--$400$ and does not saturate, which is why we do not claim a
converged continuum value (Sec.~\ref{sec:discussion}).  Panel (c): the
running estimate $\lambda_{\max}(T)$ decreases from an initial alignment
transient toward a positive asymptotic value, staying well above zero
throughout.  Panel (d): across RK4 step sizes $\Delta t=0.1,0.05,0.025$
the exponent stays far above the classification threshold
($\lambda_c=0.01$); its magnitude is not fully converged at the step used
--- it decreases toward $\sim\!0.07$ as $\Delta t\to0$ --- which is
precisely why the main text reports no $\lambda_{\max}$ magnitudes and
uses only its sign.  Panel (e): the chaos basin fraction computed from
the high-fidelity sweep is essentially unchanged as the classifier
threshold $\lambda_c$ is varied over $0.005$--$0.02$, with binomial
seed-uncertainty bands.  Thus the chaos diagnosis (a positive exponent)
and the basin cascade are robust to integration time, step size, and
classifier threshold; only the \emph{magnitude} of $\lambda_{\max}$ ---
and its $N\to\infty$ limit (panel b) --- is left unconverged, and we make
no quantitative claims about it.

\section*{Supplemental Material}

A movie is provided as Supplemental Material showing the full
spatiotemporal dynamics of the self-propelled swarmalator model at
$K=J=1$, $N=100$, sweeping through the four collective states
(traveling wave, breathing two-cluster, split wave, chaos) as $v_0$
increases.

\end{document}